\documentclass[11pt,twoside]{article}
\usepackage{hepro5}
\usepackage{graphicx}

\usepackage[T1]{fontenc} 

\usepackage{latexsym}
\usepackage{verbatim}

\newcommand{\be}{\begin{equation}}
\newcommand{\ee}{\end{equation}}
\newcommand{\bary}{\begin{eqnarray}}
\newcommand{\eary}{\end{eqnarray}}

\begin{document}

\vskip 1.0cm
\markboth{N.~Fraija}{Neutron Star Mergers}
\pagestyle{myheadings}

\vspace*{0.2cm}
\title{High-energy emission as signature of magnetic field amplification in Neutron Star Mergers}
\author{N. Fraija$^1$, W. H. Lee$^1$, P. Veres$^2$ and R. Barniol Duran$^3$}
\affil{$^1$Instituto de Astronom\'ia, Universidad Nacional Aut\'{o}noma de M\'{e}xico, Apdo. Postal 70-264, Cd. Universitaria, M\'{e}xico DF 04510.\\
$^2$ Center for Space Plasma and Aeronomic Research (CSPAR), University of Alabama in Huntsville, Huntsville, AL 35899, USA.\\ 
$^3$ Department of Physics and Astronomy, Purdue University, 525 Northwestern Avenue, West Lafayette, IN 47907, USA. ({\rm nifraija@astro.unam.mx, wlee@astro.unam.mx and pv0004@uah.edu and rbarniol@purdue.edu)}
}

\begin{abstract}
The merger of a binary neutron star system is suggested  as the central engine of short gamma-ray bursts (sGRBs).  For the merger process, simulations predict  that magnetic field is amplified beyond magnetar field strength  by Kelvin-Helmholtz instabilities.  With the Large Area Telescope (LAT), bursts have been detected that show a temporally extended component in coincidence with a short-lasting peak at the end of the prompt phase.  We show that the presence of these LAT components in a sGRB could provide evidence of magnetic field amplification in the neutron star merger.
%
\end{abstract}



\section{Introduction}
The most popular progenitor model associated with short gamma-ray bursts (sGRBs) is the merger of compact object binaries; a neutron star binary  or a neutron star - black hole (NS-BH; Eichler et al. 1989; Lee et al. 2004, 2005b).   During the merger, simulations have exhibited an amplification of magnetic field strength,  achieving values as high as $10^{15}$ G.  The growth emerges from the transfer of hydrodynamical kinetic energy to electromagnetic energy via Kevin-Helmholtz instabilities and turbulent amplification (Price \& Rosswog 2006; Giacomazzo et al. 2009).\\
In the afterglow phase, the relativistic ejecta encounters the circumburst medium generating external shocks, forward and reverse shocks.  A strong short-lived reverse shock propagates back into the ejecta, explaining the $\gamma$-ray, X-ray, optical and/or radio flares present at the end of the prompt phase (Kobayashi 2000; Kobayashi et al. 2004; Fraija et al. 2012; Fraija 2015).   The long-lived forward shock leading to a continuous softening of the afterglow spectrum describes the temporarily extended emissions in multiwavelengths (Panaitescu 2007).\\ 
GRB110731A and GRB130427A were detected from the optical to the GeV energy range (Ackermann et al., 2013; 2014). Analysis on the prompt phase at the LAT energy range ($>$ 100 MeV )  revealed a temporally extended flux lasting hundreds of seconds in coincidence with a short-lasting bright peak.  Fraija (2015) and Fraija et al. (2016a) showed that both components could be interpreted as synchrotron and SSC emissions from the forward and reverse shocks, respectively, provided that  the reverse shock evolves in a thick-shell regime and the central engines were entrained with a significant magnetic field. In this work we propose that the presence of high-energy components in sGRBs could offer evidence of the magnetic field amplification during the merger of two neutron stars.
%
%

\section{Dynamics of external Shock Model} 
In the external shock model, afterglow emission is generated when an expanding relativistic shell encounters the circumburst medium and consequently,  forward and reverse shocks are produced (Uhm \& Belodorodov 2007).  Relativistic  electrons at both shocks are accelerated and cooled down by synchrotron and Compton scattering emission. In the following subsections we will develop the dynamics of forward and reverse shocks.  We hereafter use the observable quantities: a total energy $E=5\times10^{52}$ erg,  a luminosity distance from the source  $D=3\times 10^{27}$ cm, an homogeneous density $n=0.1$ cm$^{-3}$, a duration of the GRB $T_{90}=1$ s (Berger 2014), and the values of cosmological parameters $H_0=$ 71 km s$^{-1}$ Mpc$^{-1}$, $\Omega_m=0.27$, $\Omega_\lambda=0.73$.  Also  we will use  primes (unprimes) to define the quantities in a comoving (observer) frame and the universal constants  $\mathit{c}$=$\hbar$=1 in natural units. The subscripts $\mathit{f}$ and $\mathit{r}$ refer throughout this paper to the forward and reverse shock, respectively and the convention $Q_x=Q/10^x$ will be adopted in cgs units.
\subsection{Forward shock dynamics}
The dynamics of the afterglow for a spherical ultrarelativistic and adiabatic shell propagating into a homogenous density is analyzed through the radius shock $R=\left( \frac{3}{4\pi \,m_p}\right)^{1/3} \,n^{-1/3}\,\Gamma^{-2/3} \,E^{1/3}$ and  the deceleration time $t_{dec}=\left(\frac{3}{32\,\pi\,m_p}\right)^{1/3}\,(1+z)\,n^{-1/3}\,E^{1/3}\,\Gamma^{-8/3}$, where $\Gamma$ is the bulk Lorentz factor,  $m_p$ is the proton mass and $z$ is the redshift. 
\paragraph{Synchrotron emission}
Electrons are accelerated to a power law distribution by the first order Fermi mechanism,  $N(\gamma_e)\,d\gamma_e \propto \gamma_e^{-p}\,d\gamma_e$, with $\gamma_e\geq\gamma_{m,f}=\epsilon_{e}\left(\frac {p-2} {p-1}\right) \frac {m_p} {m_e}\Gamma$  where  $m_e$ is the electron mass and $\epsilon_{e}=  U_e/(4\,\Gamma^2\,n\,m_p) $ the electron equipartition parameter defined as fraction of the energy density $U_e$ that goes to accelerate electrons.   The total energy density ($U$) is also equipartitioned to amplify the magnetic field  $\epsilon_{B,f}=U_{B,f}/U$ (with $U_{B,f}=B'^2_f/8\pi$), hence the magnetic field   can be written as $B'_f=(32\pi m_p)^{1/2}\,\epsilon^{1/2}_{B,f}\,n^{1/2}\,\Gamma\,$.
%
%
Considering the time scale  for synchrotron cooling $t_{e,syn}\simeq \frac{3m_e}{16\sigma_T}\,(1+z)\,\epsilon^{-1}_{B,f}\,n^{-1}\,\Gamma^{-3}\,\gamma_e^{-1}$ and the deceleration time, the cooling electron Lorentz factor can be written as $\gamma_{e,c,f}=\frac{3\,m_e}{16\,\sigma_T\,m_p}(1+z)\,\epsilon^{-1}_{B,f}\,n^{-1}\,\Gamma^{-3}\,t^{-1}$,   where $\sigma_T$ is the Thomson cross section. Comparing  the acceleration  $t_{acc}\simeq \frac{2\pi\,m_e}{q_e}(1+z)\,\Gamma^{-1}\,{B'}^{-1}_f \gamma_e$ and synchrotron time scales, the maximum electron Lorentz factor achieved is $\gamma_{e,max,f}=2 \left(\frac{18\pi\, q_e^2 m_p}{\sigma^2_T}\right)^{1/4}\,\epsilon^{1/4}_{B,f}\,n^{1/4}\,\Gamma^{1/2}$, with $q_e$ the elementary charge. From the electron Lorentz factors and deceleration time scales,  the synchrotron spectral breaks  are (Fraija et al. 2016a)
\bary\label{synfor_b}
E^{syn}_{\rm \gamma,m,f}&\simeq&1.19\times 10^9\,{\rm eV}  \, \left(\frac{1+z}{2}\right)^{1/2}\,\epsilon_{e}^2\,\epsilon^{1/2}_{B,f}\,E^{1/2}_{52.7}\,t^{-3/2}_0\,,\cr
E^{syn}_{\rm \gamma,c,f}&\simeq& 23.13\,{\rm eV} \, \left(\frac{1+z}{2}\right)^{-1/2}\,(1+x_f )^{-2}\,\epsilon^{-3/2}_{B,f}\,n^{-1/2}_{-1}\, E^{-1/2}_{52.7}\,t^{-1/2}_0\,,\cr
E^{syn}_{\rm \gamma,max,f}&\simeq& 180.38\times 10^9\,{\rm eV}\, \left(\frac{1+z}{2}\right)^{-5/8}\, n^{-1/8}_{-1}\,E^{1/8}_{52.7}\,t^{-3/8}_0\,.
\eary
The lightcurve in the LAT energy range generated by the ultra-relativistic electrons in the fast-cooling regime ($E^{syn}_{\rm \gamma,m,f}<E^{syn}_\gamma<E^{syn}_{\rm \gamma,max,f}$) is
\begin{equation}
\label{fcsyn_t}
(EF_{\nu})^{syn}_{\rm \gamma,f}=3.67\times 10^{-5}\,{\rm \frac{erg}{cm^2\,s}}\, \left(\frac{1+z}{2} \right)^{1.1}\,\epsilon_{B,f}^{0.1}\,\epsilon_{e}^{1.4}\,n^\frac14_{-1}\,D^{-2}_{27.3}\,E^{-0.1}_{52.7}\,t^{-1.3}_0\,\left(E^{syn}_{\gamma,8} \right)^{-0.2}\,.
\end{equation}
%
%
\subsection{Reverse shock dynamics}
For the reverse shock, a simple analytic solution can be derived taking two limiting cases, the thick- and thin-shell case, by using a critical Lorentz factor defined by 
%
\be
\Gamma_c=1.15\times 10^3\,\left(\frac{1+z}{2} \right)^{3/8}\, E^{1/8}_{52.7}\,n^{-1/8}_{-1}\,T_{90,0}^{-3/8}\,.
\ee
For $\Gamma>\Gamma_c$ (thick shell) the shell is significantly decelerated by the reverse shock, otherwise, $\Gamma<\Gamma_c$ (thin shell), the reverse shock cannot decelerate the shell effectively.    Irrespective of the evolution of the reverse shock,  the synchrotron  spectral evolution between reverse and forward shock is related by $E^{syn}_{\rm \gamma, m,r}\sim \mathcal{R}^{-1}_B\,\mathcal{R}^{-2}_M\,E^{syn}_{\rm \gamma,m,f}$,  $E^{syn}_{\rm \gamma,c,r}\sim\,\mathcal{R}^{3}_B\,E^{syn}_{\rm \gamma,c,f}$ and $F^{syn}_{\rm \gamma,max,r}\sim\,\mathcal{R}^{-1}_B\,\mathcal{R}_M\,F^{syn}_{\gamma,max,f}$, where $\mathcal{R}_B=\frac{B'_f}{B'_r},\hspace{0.2cm} {\rm and} \hspace{0.2cm} \mathcal{R}_M=\frac{\Gamma^2_{d}}{\Gamma}$ with $\Gamma_{d}$ the bulk Lorentz factor at the shock crossing time $t_d$.\\ 
%
%
%
\noindent  
In the thick-shell regime,  the reverse shock becomes relativistic during its propagation and the ejecta is significantly decelerated.  The bulk Lorentz factor at the shock crossing time  $t_d\simeq T_{90}$ is given by $\Gamma_{d}\sim \Gamma_c$.  Eventually,  the shock crossing time could be shorter than $T_{90}$ depending on the degree of magnetization of the ejecta (Zhang \& Kobayashi 2005).  Numerical analysis performed by Fan et al. (2004) revealed that for the particular value of the magnetization parameter $\sigma \simeq 1$, the shock crossing time becomes $t_d\simeq T_{90}/6$.\\
\paragraph{Synchrotron emission.}
Assuming that electrons are accelerated in the reverse shock to a power-law distribution and the energy density is equipartitioned  between electrons and magnetic field, then the minimum electron Lorentz factor $\gamma_{\rm e,m,r}=\epsilon_{e}\left(\frac {p-2} {p-1}\right) \frac {m_p} {m_e}\frac{\Gamma}{\Gamma_{d}}$ and magnetic field is 
\be
B'_r=(32\pi m_p)^{1/2}\,\epsilon^{1/2}_{B,r}\,n^{1/2}\,\Gamma\,.
\ee
From the characteristic cooling time of synchrotron radiation and dynamical time scale, the cooling Lorentz factor is  $\gamma_{e,c,r}=\frac{3\,m_e}{16\,\sigma_T\,m_p}(1+z)\,(1+x_r)^{-1}\,\epsilon^{-1}_{B,r}\,n^{-1}\,\Gamma^{-3}\,T_{90}^{-1}$.  Additionally,  from  synchrotron break energy relations (forward and reverse shock), we can write the synchrotron spectral breaks as 
%
%
\bary\label{synrev_b}
E^{syn}_{\rm \gamma,m,r}&\sim& 6.94\times 10^{-4}\, {\rm eV}\,\left(\frac{1+z}{2}\right)^{-1}\,\epsilon_{e}^{2}\,\epsilon_{B,r}^{1/2}\,\Gamma^{2}\,n^{1/2}_{-1}, \cr
E^{syn}_{\rm \gamma,c,r}&\sim& 2.58\, {\rm eV}  \,\left(\frac{1+z}{2}\right)^{-1/2}\,(1+x_r)^{-2}\,\epsilon_{B,r}^{-3/2}\,n^{-1}_{-1}\,E^{-1/2}_{52.7}\,T_{90,0}^{-1/2}\,.
\eary
Here, $x_r=\frac{-1+\sqrt{1+4\eta \epsilon_{e}/\epsilon_{B,r}}}{2}$ is  the ratio of the SSC to synchrotron luminosity, where $\eta$ for slow-cooling and fast-cooling regime is  $(\gamma_{\rm e,c,r}/\gamma_{\rm e,m,r})^{2-p}$  and  $\eta=1$, respectively (Sari \& Esin 2001).  
\paragraph{SSC emission.}
Accelerated electrons can upscatter photons from low to high energies as $E^{ssc}_{\gamma,m,r}\sim\gamma^2_{\rm e,m,r},E_{\gamma,m,r}$,  $E^{ssc}_{\rm \gamma,c}\sim\gamma^2_{\rm e,c,r}\,E_{c,r}$ and $F^{ssc}_{\rm\gamma,max,r}\sim\,k\tau\,F_{\rm max,r}$ where $k=4(p-1)/(p-2)$ and $\tau=\frac{\sigma_T N(\gamma_e)}{4\pi r_d^2}$ is the optical depth of the shell.  From electron Lorentz factors and the synchrotron spectral breaks (eq. \ref{synrev_b}),  we get the break SSC energies     
\bary\label{ssc_b}
E^{ssc}_{\rm \gamma,m,r}&\sim& 5.83\times 10^{-4}\, {\rm eV}   \,\left(\frac{1+z}{2}\right)^{-7/4}\,\epsilon_{e}^{4}\,\epsilon_{B,r}^{1/2}\,\Gamma^{4}\,n^{3/4}_{-1}\,E^{-1/4}_{52.7}\,T_{90,0}^{3/4}\,,\cr
E^{ssc}_{\rm \gamma,c,r}&\sim& 9.44\times 10^{22}\, {\rm eV}  \,\left(\frac{1+z}{2}\right)^{3/2}\,(1+x_r)^{-4}\,\epsilon_{B,r}^{-7/2}\,n^{-3}_{-1}\,E^{-1/2}_{52.7}\,\Gamma^{-6}\,T_{90,0}^{-5/2}\,.
\eary
%
%
%
Considering the SSC spectrum in the fast-cooling regime,  the SSC flux reaches the peak $F^{ssc}_{\rm \gamma,peak,r}\sim (E^{ssc}_{\rm \gamma,r}/ E^{ssc}_{\rm \gamma,c,r})^{-1/2}\,F^{ssc}_{\rm \gamma,max,r}$ at (Kobayashi \& Zhang 2003):
\bary\label{ssc_peak}
(EF_\nu)^{ssc}_{\rm \gamma,peak}&\sim& 1.25\times 10^{11}\,{\rm \frac{erg}{cm^2\,s}}\,(1+x_r)^{-2}\,\left(\frac{1+z}{2}\right)^3\,\epsilon_{B,r}^{-3/4}n^{-3/4}_{-1}D^{-2}_{27.3}\,E^{3/2}_{52.7}\,\cr
&&\hspace{5.5cm}\times\, \Gamma^{-5}\,T_{90,0}^{-5/2}\,(E^{ssc}_{\gamma,r,8})^{1/2}.
\eary
%
%
%
\section{Results and Conclusions}
\begin{figure} 
\includegraphics[height=7.6cm]{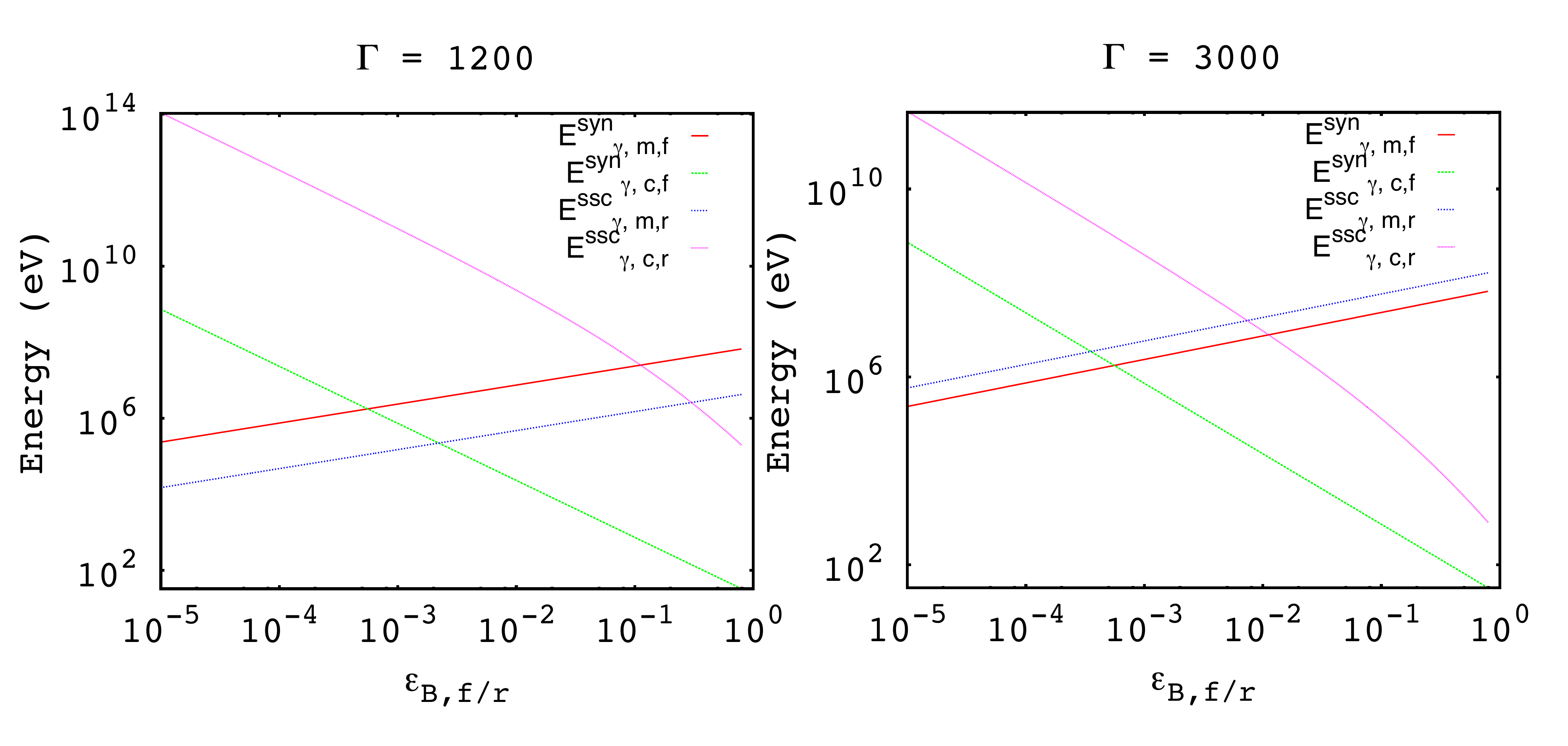}
\caption{$E^{syn/ssc}_{\rm \gamma,f/r}$ vs  $\epsilon_{B,f/r}$ for $\Gamma=1200$ (left) and $\Gamma=3000$ (right).}
\label{fig1}
\end{figure}

\noindent We plot the synchrotron and SSC spectral breaks of the forward and reverse shocks as a function of equipartition parameters ($\epsilon_{B,f/r}$ and $\epsilon_{e}$), considering the typical values of the magnetic  ($10^{-5}\,\leq\epsilon_{B,f(r)}\leq\,1$) and electron ($\epsilon_{e,f}$= 0.25) parameters (Santana et al. 2014) for $\Gamma=1200$ (left panel) and $\Gamma=3000$ (right panel), as shown in Figure \ref{fig1}.
%
%
%
%
When we consider $\Gamma=1200$,  the break energies behave as $E^{syn}_{\rm \gamma,m,f} <  E^{syn}_{\rm \gamma,c,f}$ ($E^{syn}_{\rm \gamma,c,f}  < E^{syn}_{\rm \gamma,m,f}$) and  $E^{ssc}_{\rm \gamma,m,r} <  E^{ssc}_{\rm \gamma,c,r}$ ($E^{ssc}_{\rm \gamma,c,r} < E^{ssc}_{\rm \gamma,m,r}$) for  $\epsilon_{B,f/r}\sim 10^{-5}$ ($\epsilon_{B,f/r}> 0.5$) and  when $\Gamma=3000$ is assumed, a shift of parameter ($\epsilon_{B,f/r}> 0.01$) is obtained for the break energies  $E^{syn}_{\rm \gamma,c,f}  < E^{syn}_{\rm \gamma,m,f}$ and $E^{ssc}_{\rm \gamma,c,r} < E^{ssc}_{\rm \gamma,m,r}$. Therefore, for small values of $\epsilon_B$, the synchrotron and SSC emissions are at slow-cooling regime whereas for large values of $\epsilon_B$ both emissions are at fast-cooling regime. Also when the bulk Lorentz factor increases (from $\Gamma=1200$ to  $\Gamma=3000$) the characteristic break energies change (from $E^{ssc}_{\rm \gamma,m,r}<E^{syn}_{\rm \gamma,m,r}$ to $E^{ssc}_{\rm \gamma,m,r}>E^{syn}_{\rm \gamma,m,r}$) for any value of $\epsilon_B$.\\
Using the method of Chi-square $\chi^2$ minimization as implemented in the ROOT software package (Brun \& Rademakers 1987), we obtain the values of parameters $\epsilon_{B,f(r)}$ and $\epsilon_{e}$ that could describe the temporally extended flux superposed with the short-lasting bright peak at the LAT energy range, considering a typical LAT flux density (from $\sim 10^{-6}$  to  $\sim10^{-5}\, {\rm \frac{erg}{cm^2\,s}}$ at 100 MeV) and a spectral index of electron distribution $p=2.4$ reported in most of the bursts detected by Fermi-LAT (Ackermann et al. 2010, 2011, 2013). The temporally extended flux is explained by synchrotron radiation from forward shock and the short-lasting bright peak with SSC emission from the reverse shock.\\ 
\begin{figure} 
\includegraphics[height=7.0cm]{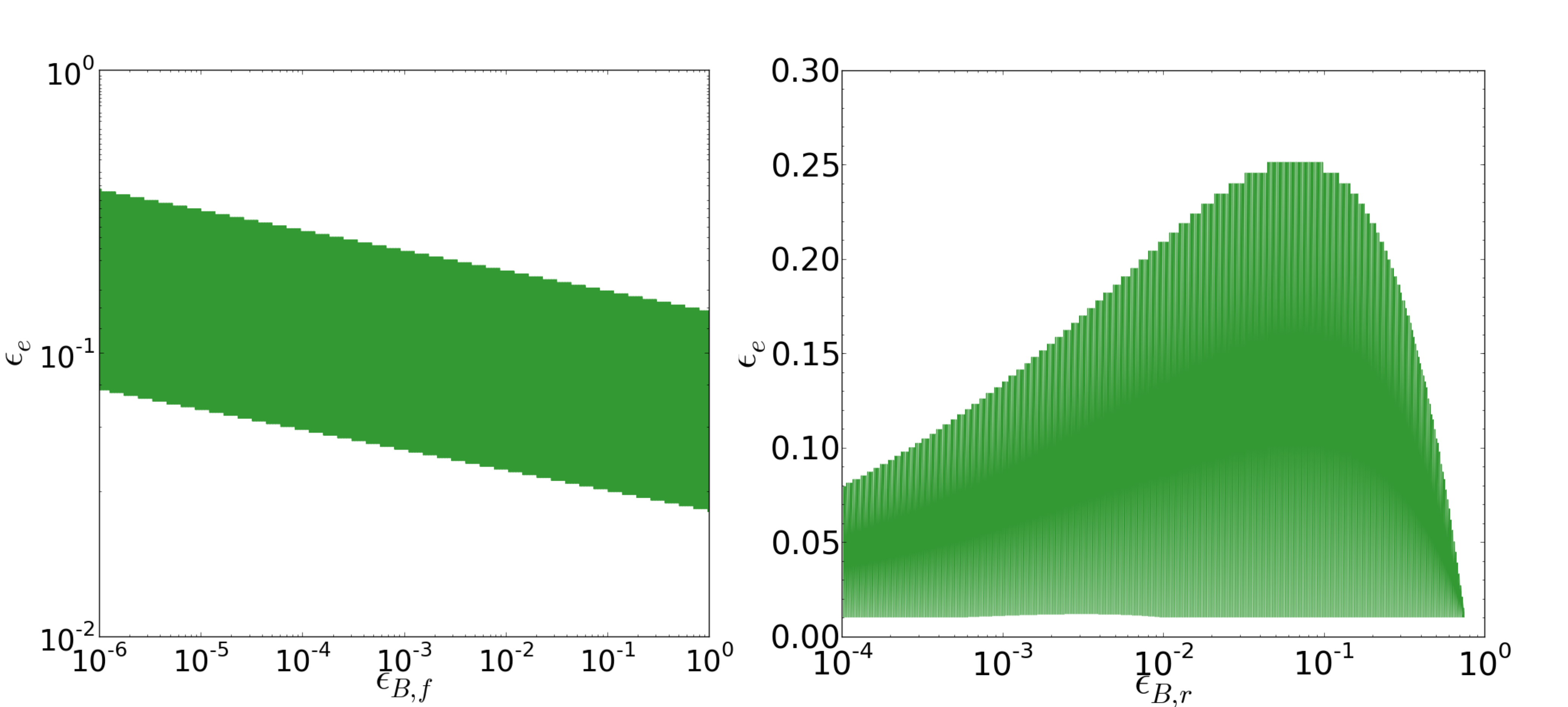}
\caption{Values of $\epsilon_e$ and $\epsilon_{B,f/r}$ that explain the temporarily extended (left)  and the short-peak bright (right)  LAT  flux}
\label{fig2}
\end{figure}

%
We can see that the difference between the synchrotron and SSC spectral breaks achieved in forward and reverse shocks, respectively,  are explained through the energy distribution given to magnetic field.  Comparing the magnetic equipartition parameters that best describe the emission at forward and reverse shocks, we can see that magnetic fields in both shocks are related by  $B_f\simeq 10^{-1}\,B_r$. The previous result suggests that the magnetic field in the reverse-shock region is stronger ($\sim$ 10 times) than in the forward-shock region,  indicating that the ejecta is magnetized and hence, the magnetic field amplification during the merger of NS - NS.

\end{document}